\documentclass[aps,prl,twocolumn,nofootinbib,superscriptaddress]{revtex4}%
\usepackage{color}
\usepackage{epsfig}
\usepackage{graphicx}
\usepackage{amsmath}
\usepackage{amsfonts}
\usepackage{amssymb}
\usepackage{epstopdf}
\usepackage{bbold}%
\providecommand{\U}[1]{\protect\rule{.1in}{.1in}}
\def\be{\begin{equation}}
\def\ee{\end{equation}}
\begin{document}
\title{Super Yang-Mills Theory as a Twistor Matrix Model}
\author{Jonathan J. Heckman}
\affiliation{School of Natural Sciences, Institute for Advanced Study, Princeton, NJ 08540}
\author{Herman Verlinde}
\affiliation{Department of Physics, Princeton University, Princeton, NJ 08544}

\begin{abstract}
We introduce a covariant finite regulator for $\mathcal{N} = 4$ super
Yang-Mills theory on $S^{4}$. Our formulation is based on holomorphic
Chern-Simons theory on twistor space. By switching on a large background flux,
the twistor space dissolves into a fuzzy geometry, with a finite number of
points. The large $N$ continuum limit of the matrix model naturally
approaches ordinary $\mathcal{N}\! =\!4$ SYM.
We comment on the relation between our model
and the 4D quantum Hall effect.

\end{abstract}
\maketitle

\vspace{-2mm}

\subsection{{\protect\normalsize Introduction}}

\vspace{-3mm}

Considerations from holography suggest that quantum field theory involves an overcomplete number of degrees
of freedom. It is believed that string theory somehow avoids this
overcounting, as evidenced by the existence of dual formulations such as
AdS/CFT. Even so, the microscopic origin of holography remains poorly
understood, especially in the context of space-times with positive curvature.
One way to try to make some progress with this question is to look for
self-consistent covariant ways of regulating gauge theories with holographic duals.

In this short note, we present a concrete proposal for a covariant regulator
of $U(N_{c})$ $\mathcal{N}$\,= 4 super Yang-Mills theory (SYM) on $S^{4}$ in
terms of a non-commutative, or ``fuzzy'' twistor space with a finite number of fuzzy points.
The theory is obtained via a decoupling limit from ordinary holomorphic Chern-Simons (hCS) theory on
super-twistor space with a large background flux switched on. The presence of
the flux fuzzifies the space, and also generates a $BC$ system localized on
$D6/D2$-brane intersections wrapping the twistor lines. Via the twistor correspondence,
the large $N$ continuum limit of the matrix model naturally reproduces ordinary $\mathcal{N}\! =\!
4$ SYM theory on $S^{4}$.

\vspace{-3mm}

\subsection{{\protect\normalsize Twistor Gauge Theory}}

\vspace{-3mm} Twistors (see \cite{PenroseRindler} for a review) provide a
convenient way to characterize massless theories. In complexified Minkowski
space $\mathbb{M}$, light-like momenta $p_{AA^{\prime}}$ can
be represented as a product $p_{AA^{\prime}}=\widetilde{\pi}_{A}%
\pi_{A^{\prime}}$ of a chiral and an anti-chiral spinor. Given
a space-time point $x^{AA^{\prime}}$, we can define a corresponding
spinor $\omega^{A}=ix^{AA^{\prime}}\pi_{A^{\prime}}$, and thereby obtain a set
of homogeneous coordinates $Z^{\alpha}=(\omega^{A},\pi_{A^{\prime}})$ on
projective twistor space $\mathbb{PT}^{\bullet}$. The dual twistor space, with
coordinates $\widetilde{Z}_{\beta}=(\widetilde{\pi}_{A},\widetilde{\omega
}^{A^{\prime}})$, is denoted by $\mathbb{PT}_{\bullet}$. Geometrically,
$\mathbb{PT}^{\bullet}$ and its dual both define a $\mathbb{CP}^{3}$.

A basic feature of twistor theory is that physical space-time is derived from
twistor data. The two equations
\begin{equation}
\omega^{A}=ix^{AA^{\prime}}\pi_{A^{\prime}}%
\end{equation}
for {\small $A$=1,2} cut out a $\mathbb{CP}_{x}^{1}$ in $\mathbb{PT}^{\bullet
}$, which is the twistor line associated with a single point $x^{AA^{\prime}}$
in $\mathbb{M}$. Conversely, a point in twistor space corresponds to a
null-plane in $\mathbb{M}$.

This correspondence can be extended to superspace coordinates $\mathcal{X}%
^{II^{\prime}}=(x^{AA^{\prime}}|\theta^{A^{\prime} i})$, and supertwistor coordinates
$\mathcal{Z}^{I}=(\omega^{A},\pi_{A^{\prime}}|\psi^{i})$. The resulting
supertwistor space $\mathbb{CP}^{3|4}$ is a Calabi-Yau manifold, a
natural target space for the topological string \cite{Witten:2003nn}. The open
string sector of the B-model is holomorphic Chern-Simons theory
\begin{equation}
S_{hCS}=\int_{\strut\strut\strut\mathbb{CP}^{3|4}}\!\!\!\Omega\wedge
\mathrm{Tr}\Bigl(\mathcal{A}\overline{\partial}\mathcal{A}+{\textstyle\frac
{2}{3}}\mathcal{A}^{3}\Bigr) \label{hCS}%
\end{equation}
where $\mathcal{A}=\mathcal{A}_{I}d\overline{\mathcal{Z}}^{I}$ denotes a gauge
superfield. This theory is physically equivalent to self-dual $\mathcal{N}\!=\!4$
SYM \cite{Witten:2003nn}, provided that ${\mathcal{A}}$ is
restricted to be flat over each twistor line. Each component of the holomorphic superfield $\mathcal{A}_{I}$
corresponds to a field of the 4D SYM theory.

\vspace{-3mm}

\subsection{{\protect\normalsize Twistor Quantization}}

\vspace{-3mm}

The product $\mathbb{PT}^{\bullet} \times \mathbb{PT}_{\bullet}$ is
naturally viewed as a quantized phase space
\cite{Penrose:1968me}. The twistor commutation algebra
\begin{equation}
\label{zcom}\lbrack Z^{\alpha},\widetilde{Z}_{\beta}]=\hslash\delta_{\beta
}^{\alpha}\Longrightarrow\left\{
\begin{array}
[c]{l}%
\lbrack\omega^{A},\widetilde{\pi}_{B}]=\hslash\delta_{B}^{A}\\[1mm]%
\lbrack\pi_{A^{\prime}},\widetilde{\omega}^{B^{\prime}}]=\hslash
\delta_{A^{\prime}}^{B^{\prime}}%
\end{array}
\right\}  .
\end{equation}
provides a representation of $sl(4,\mathbb{C})$, the algebra of
the complexified conformal group with generators:%
\begin{align}
\textstyle P_{AA^{\prime}}  &  =\widetilde{\pi}_{A}\pi_{A^{\prime}%
}\ \ ,\ \ K_{AA^{\prime}}=\widetilde{\omega}_{A^{\prime}}\omega_{A}%
\,,\nonumber\\[.5mm]
J_{AB}  &  =\widetilde{\pi}_{(A}\omega_{B)}\ ,\ \text{ }\widetilde
{J}_{A^{\prime}B^{\prime}}=\widetilde{\omega}_{(A^{\prime}}\pi_{B^{\prime}%
)}\,,\\[.5mm]
D  &  ={\textstyle\frac{1}{2}}\bigl(\widetilde{\pi}^{A}\omega_{A}%
-\widetilde{\omega}_{B^{\prime}}\pi^{B^{\prime}}\bigr).\nonumber
\end{align}
The extension to space-time supersymmetric theories is achieved by including
four fermionic oscillators $\psi^{i}$ with $\bigl\{\psi^{i},\widetilde{\psi
}_{j}\bigr\}=\hslash\delta_{j}^{i}$, which then provides a representation of
$sl(4|4)$. In what follows we shall often leave implicit the extension to the
fermionic case.

Specifying a real space-time signature amounts to a choice of realization of
$sl(4,\mathbb{C})$, \textit{i.e.} a convention for Hermitian conjugation. The
three signatures $(++++)$, $(++--)$ and $(-+++)$ correspond to the
realizations of $sl(4,%
\mathbb{C}
)$ respectively by $su(4)$, $sl(4,%
\mathbb{R}
)$ and~$su(2,2)$. Explicit realizations of each case are:%
\begin{align}
su(4)  &  :\ \ \ \omega_{A}^{\dag}=\widetilde{\pi}_{A}\text{, }\pi_{B^{\prime
}}^{\dag}=\widetilde{\omega}_{B^{\prime}}\text{\ \ \ and\ \ \ }\hslash\in%
\mathbb{R}%
\label{su4def}\\[1mm]
su(2,2)  &  :\ \ \ \omega_{A}^{\dag}=\widetilde{\omega}_{A^{\prime}}\text{,
}\pi_{A^{\prime}}^{\dag}=\widetilde{\pi}_{A}\text{\ \ \ and\ \ \ }\hslash\in%
\mathbb{R}%
\\
sl(4,%
\mathbb{R}
)  &  :\left\{
\begin{array}
[c]{c}%
\omega_{A}^{\dag}=\omega_{A}\text{, }\pi_{A^{\prime}}^{\dag}=\pi_{A^{\prime}%
}\\[1mm]%
\widetilde{\omega}_{A^{\prime}}^{\dag}=\widetilde{\omega}_{A^{\prime}}\text{,
}\widetilde{\pi}_{A^{\prime}}^{\dag}=\widetilde{\pi}_{A^{\prime}}%
\end{array}
\right\}  \text{ and }\hslash\in i%
\mathbb{R}%
\end{align}
The reality condition on $\hslash$ is enforced by Hermitian conjugation on the
commutators.

In the remainder of this note we restrict to Euclidean signature. The
hermitian conjugation map (\ref{su4def}) provides an identification
$(\mathbb{PT}_{\bullet})^{\dag} = \mathbb{PT}^{\bullet}$ which acts on the conformal symmetry
generators as:%
\begin{equation}
P^{\dag}=K\text{, }J^{\dag}=J\text{, }\widetilde{J}^{\dag}=\widetilde
{J}\text{, }D^{\dag}=-D\text{,}%
\end{equation}
leaving the $SO(5)$ generators $J$, $\widetilde{J}$ and $\frac{1}{2}(P+K)$ as
the only hermitian charges. The Euclidean theory is
therefore most naturally viewed as a radially quantized theory on $S^{4}$.
The hermitian charges for the supersymmetric case
are the symmetries of the supersphere $S^{4|8}$.

The projection $\mathbb{CP}^{3}\rightarrow S^{4}$ is achieved by introducing the
quaternionic coordinates
\begin{equation}
\label{quat}(Q_{1},Q_{2}) = (Z_{1}+j Z_{2},Z_{3}+jZ_{4})\in\mathbb{HP}%
^{1}=S^{4}%
\end{equation}
with $\mathbb{HP}^{1}$ the quaternionic projective plane, and $j$ a
quaternion. We will make important use of this quaternionic perspective momentarily.

We see that the commutator algebra (\ref{zcom}) naturally identifies the
twistor coordinates with creation and annihilation operators, and functions on
$\mathbb{PT}$ with linear operators on the associated Fock space. Introduce a
vacuum state
such that $Z^{\alpha}\left\vert \text{vac}\right\rangle \! =\! \psi
^{i}\left\vert \text{vac}\right\rangle \!=\! 0$. The Fock space is:%
\begin{equation}
\mathcal{F}=\text{span}_{%
\mathbb{C}
}\Bigl\{\,\underset{\beta,i}{%
{\displaystyle\prod}
}\,{\widetilde{Z}_{\beta}^{n_{\beta}}}\widetilde{{\psi}}_{i}^{n_{i}}\left\vert
\text{vac}\right\rangle \Bigr\}.
\end{equation}
This space is graded by the homogeneity operator:
\begin{equation}\label{DTERM}
\mathfrak{D}=\widetilde{Z}_{\beta}Z^{\beta}+\widetilde{\psi}_{i}\psi^{i}.
\end{equation}
The level $N$ subspace
\begin{equation}
\mathfrak{D}\left\vert \Psi\right\rangle = \hbar N\left\vert \Psi\right\rangle
\end{equation}
describes the Hilbert space $\mathcal{H}_{\mathbb{PT}}(N)$ of fuzzy points on
a non-commutative $\mathbb{CP}^{3|4}$ with K\"ahler parameter $\hslash N$. The bosonic
subspace of $\mathcal{H}_{\mathbb{PT}}(N)$ consists of the $N$-fold symmetric
product of the fundamental of $sl(4,\mathbb{C})$, and thus fills out an
irreducible representation of dimension
\begin{equation}
\label{kfuzz}k = (N+1)(N+2)(N+3)/6.
\end{equation}

The analogue of the twistor equation in $\mathcal{H}_{\mathbb{PT}}$ is%
\begin{equation}
\label{line}(\omega^{A}-ix^{AA^{\prime}}\pi_{A^{\prime}})\left\vert
q_{x}\right\rangle =0,
\end{equation}
where $x^{AA^{\prime}}$ is a complex $2\times2$ matrix defining a point in
complexified space-time. Eq. (\ref{line}) projects onto a fuzzy $\mathbb{CP}%
^{1}$ inside $\mathbb{CP}^{3}$. We refer to this subspace as $\mathcal{H}%
_{x}\subset\mathcal{H}_{\mathbb{PT}}$. Each subspace $\mathcal{H}_{x}$ has dimension $N+1$ and transforms as a spin
$j=N/2$ representation of $sl(2,\mathbb{C})$.

Note that the collection of all position eigenstates $\left\vert \left\vert
x\right)  \right)  $ forms an overcomplete basis. Pick a state $\left\vert
q_{x}\right\rangle \in\mathcal{H}_{x}$ and an $SO(5)$ rotation $R_{y,x}$ which
maps $x$ to $y$. We can compute the overlap $\left(  \left(  y||x\right)
\right)  =\left\langle q_{x}\right\vert R_{y,x}\left\vert q_{x}\right\rangle
$.
In terms of the rotation angle $\theta_{x,y}$ from $x$ to $y$
\begin{equation}
\label{overlap}
\left(  \left(  y||x\right)  \right)  =\bigl(\cos{\textstyle\frac{1}{2}}%
\theta_{x,y}\bigr)^{N}.
\end{equation}
At large $N$ this is a sharply peaked Gaussian function of width
$\sigma=2/\sqrt{N}$.

The Hilbert space $\mathcal{H}_{\mathbb{PT}}$ is closely related to the
realization of the quantum Hall effect (QHE) on $S^{4}$ given in
\cite{Zhang:2001xs}. In particular, the dimension of $\mathcal{H}_{\mathbb{PT}}$ matches with that
of the lowest Landau level (LLL) found in \cite{Zhang:2001xs}.

\subsection{{\protect\normalsize A Decoupling Limit}}

\vspace{-3mm}

To obtain the twistor matrix model, we start with holomorphic Chern-Simons
theory on commutative super twistor space. The idea is to turn on a suitable gauge bundle with a
large instanton charge and isolate the effective theory that describes the
lightest fluctuations around this background. Motivated by the correspondence
with the QHE \cite{Zhang:2001xs}, the gauge field we choose to turn on is a
lift of the Yang monopole \cite{Yang:1977qv}.  As we will see, this procedure leads to
an hCS theory defined on fuzzy twistor space, coupled to fundamental matter
localized on the twistor lines. In what follows, we will view the hCS theory as a decoupled
subsector of an open string theory.

Consider hCS theory with gauge group $G=U(n N_{c})$ where $n\equiv N+1$. We
switch on a $U(1)\times SU(n)$ gauge field ${\mathcal{A}}_{Y}$ which breaks
the gauge group $G$ to $U(N_{c})$. $\mathcal{A}_{Y}$ is obtained from the Yang
monopole
\begin{equation}
{\mathcal{A}}_{Y}=\textstyle\frac{1}{2}(dQ_{a}^{\dag}Q_{a}-Q_{a}^{\dag}dQ_{a})
\end{equation}
with $Q_{a}$ defined in (\ref{quat}), via the replacement \cite{Zhang:2001xs}
\begin{equation}
1\rightarrow N\mathbb{1}\ \quad i\rightarrow -2i \mathbf{I}_{1}\quad
j\rightarrow -2i \mathbf{I}_{2}\quad k\rightarrow -2i \mathbf{I}_{3}%
\end{equation}
with $\mathbb{1}$ the $U(1)$ generator and $\mathbf{I}_{a}$ the spin $N/2$
generators of $SU(2)$. The non-abelian part of ${\mathcal{A}}_{Y}$ is trivial
along the $\mathbb{CP}^{1}$ fibers, and maps via the twistor correspondence to
a homogeneous $SU(n)$ instanton on $S^{4}$ with instanton number
$k_{\mathrm{inst}}\!=n(n^{2}-1)/6$. The abelian component of $\mathcal{A}_{Y}$
is non-trivial along the $\mathbb{CP}^{1}$ fiber direction: each twistor line
carries $N$ units of $U(1)$ flux.

Let us consider the low energy fluctuations of the hCS theory around this
background. To organize the low energy content, it is helpful to
translate the bundle data into the language of D-branes. To avoid clutter, we
take $N_{c}\!=\! 1$; the generalization to $N_{c}\! >1$ is straightforward.

Twistor space contains $\mathbb{CP}^{p}$ for $p\leq3$ as a $2p$-cycle. Brane
bound states are therefore labeled by charge vectors $Q\!=\!(q_{6},q_{4}%
,q_{2},q_{0})$ with $q_{p}$
the $Dp$-brane wrapping number around the $p$-cycle. $U(n)$ hCS theory in the
$\mathcal{A}_{Y}$ background can be viewed as the world volume theory on a
bound state of $n$ space-filling $D6$-branes with $k$ $D2$-branes wrapping
the~$\mathbb{CP}^{1}$, with $k$ given by eq. (\ref{kfuzz}). In addition, the non-zero $U(1)$ flux through
$\mathbb{CP}^{1}$ induces on each $D6$ an opposite $D4$ charge, while each $D2$
acquires an opposite $D0$ charge. The total brane configuration has a charge
vector $Q = (n,-n,-k,k)$, and naturally splits up as an $n$-stack of
$D6/D4$-branes with charge vector $(n,-n,0,0)$ and a $k$-stack of
$D2/D0$-branes with charge vector $(0,0,-k,k)$. The open string ground states
thus organize into matrices of four different sizes
\begin{equation}
\label{xbcy}X_{k \times k}, \; B_{n\times k},\; C_{k\times n},\; Y_{n\times n}%
\end{equation}
$X$ describes the collective motion of the $D2/D0$-branes, and $Y$ denotes the
gauge and adjoint matter fields on the $D6/D4$-branes. $B$ and $C$ are the open
strings that live on the intersections between the two stacks of branes.

An important consequence of the extra $U(1)$ flux is that it lifts the
degeneracy of the ADHM moduli space. Let $V_{6}$ and $V_{2}$ denote the
bundles associated with the $D6/D4$ $n$-stack and $D2/D0$ $k$-stack. We can count
the deformations of the bound state bundle via
\begin{align}
\chi(V_{6},V_{2}) =  &  \int\mathrm{ch}(V^{\vee}_{6})\mathrm{ch}(V_{2})\mathrm{Td}(\mathbb{CP}^{3})\nonumber\\[1mm]
=  &\;  2kn - kn - kn\; =\, 0,\nonumber
\end{align}
where we used that $\mathrm{Td}(\mathbb{CP}^{3}) = 1+ 2H + \ldots$ with $H$
the hyperplane class divisor, representing the $\mathbb{CP}^{2}$ 4-cycle. The
physical mechanism that lifts the moduli is that in terms of the
$U(n)$ gauge theory on the $D6/D4$-brane, the $U(1)$ flux represents a
non-trivial vev for the adjoint scalars. This moves the $D6/D4$-brane gauge theory
slightly onto its Myers branch. The ADHM moduli still survive as
the relevant low energy degrees of freedom, but they are no longer exact zero modes.

\vspace{-3mm}

\subsection{{\protect\normalsize Twistor Matrix Model}}

\vspace{-3mm}

Let us summarize our proposal. The effective theory of the light fluctuations
of the $U(n N_{c})$ hCS theory in the ${\mathcal{A}}_{Y}$ background, we
propose, takes the form of a $U(N_{c})$ hCS theory on a fuzzy $\mathbb{CP}%
^{3|4}$ with a $k$ dimensional Hilbert space $\mathcal{H}_{\mathbb{PT}}(N)$
with $n\!=\!N\!+1$. In addition, the theory contains bifundamentals $B$ and
$C$ connecting the $k$ stack of $D2/D0$-branes to the $n$ stack of $D6/D4$-branes.
From the open string perspective, the fuzzification arises
because the open string end points are charged under the background flux.
The presence of the flux introduces an explicit
choice of scale via $\mathrm{{Vol}(\mathbb{PT})=\hslash^{3}\dim
}\mathcal{H}\mathrm{_{\mathbb{PT}}}$, and the open string spectrum is
gapped, with excitations of energy ${\hslash}^{-1/2} \sim N^{1/2} \times \mathrm{{Vol}}(\mathbb{PT})^{-1/6}$. The excitations
decouple at large $N$, and in the low energy limit the modes are forced to lie in the LLL.
We will now construct the action for this low energy effective theory
(see also \cite{Lechtenfeld:2005xi}).

Given a commutative theory on a toric space, there exists a natural way to
fuzzify the theory, while leaving the holomorphic geometry intact
\cite{Heckman:2010pv}. Here we apply this technology to the twistor gauge theory.

The hCS sector on fuzzy $\mathbb{PT}$ captures the collective dynamics of the
$D0$-branes, described by the $k\!\times\! k$ matrix $X$ in eq.~(\ref{xbcy}).
It consists of a $(0,1)$-form superfield $\mathcal{A=A}_{I}(Z^{\dag}%
,Z,\psi^{\dag},\psi)d\overline{\mathcal{Z}}{}^{I}$, subject to the homogeneity
constraint $[\mathfrak{D},\mathcal{A}_{I}]=0$, with $\frak{D}$ given in eq. (\ref{DTERM}).

Given an element $h(Z^{\dag},Z,\psi^{\dag},\psi)$ with coefficients in the Lie
algebra $u(N_{c})$, a gauge transformation on $\mathcal{A}_{I}$ is given by
$\mathcal{A}\rightarrow e^{-h}\!\ast\!\mathcal{A\,\ast}\,e^{h}+e^{-h}%
\!\ast\overline{\partial}e^{h}$. Here $\overline{\partial}$ is the Dolbeault
operator on fuzzy $%
\mathbb{C}
^{4|4}$, which acts by commutation with the holomorphic coordinates.
The holomorphic Chern-Simons functional is $hCS(\mathcal{A})=$
Tr$_{u(N_{c})}\left(  \mathcal{A}\wedge_{\ast}\overline{\partial}%
\mathcal{A}+\frac{2}{3}\mathcal{A}\wedge_{\ast}\mathcal{A}\wedge_{\ast
}\mathcal{A}\right)  $. The action is then:%
\begin{equation}\label{gmm}
\mathcal{S}_{hCS}=\frac{1}{g_{\mathrm{MM}}^{2}}\text{Tr}_{\mathcal{H}_{\mathbb{PT}}%
}\bigl(\Omega\wedge_{\ast}hCS(\mathcal{A})\bigr)
\end{equation}
where $\Omega=\frac{\varepsilon_{IJKL}}{4!}Z^{I}dZ^{J}dZ^{K}dZ^{L}%
\frac{\varepsilon^{ijkl}}{4!}\psi_{i}^{\dag}\psi_{j}^{\dag}\psi_{k}^{\dag}%
\psi_{l}^{\dag}$ and the trace over $\mathcal{H}_{\mathbb{PT}}$ is over all
fuzzy points of the superspace consistent with the homogeneity constraint. The
presence of the $\psi^{\dag}$'s in the measure factor ensures the
analogue of the relation $\int d^{2}\psi\overline{\psi}f(\psi)=\int d\psi
f(\psi)$.

The $B$ and $C$ modes are (transposes of) linear maps $\mathcal{H}%
_{\mathbb{PT}}\rightarrow\mathcal{H}_{\mathbb{CP}^{1}}$. Geometrically, we can view these modes as
localized on $D2/D6$-brane intersections along the twistor lines. Just as for
other intersecting brane configurations with localized modes, we treat them as
bulk $(0,1)$ forms with support along an appropriate subspace. The interaction
terms between the gauge field $\mathcal{A}$ and the $B$ and $C$ modes are
found by expanding a parent $U(k+n)$ hCS theory around a breaking pattern to
$U(k)\times U(n)$. The off-diagonal modes are the $B$, $C$ fields, with
action:%
\begin{equation}
\mathcal{S}_{BC}=\text{Tr}_{\mathcal{H}_{\mathbb{CP}^1}}\bigl(B\cdot \overline{D}_{\mathbb{PT}}\cdot
C\bigl) - \text{Tr}_{\mathcal{H}_{\mathbb{PT}}}\bigl(C\cdot \overline{\partial}_{\mathbb{CP}^{1}}\cdot B\bigr). \label{SBC}%
\end{equation}
Here, $\overline{D}_{\mathbb{PT}}=\Omega\wedge_{\ast}(\overline{\partial
}_{\mathbb{PT}}+\mathcal{A})$ is the covariant derivative acting by matrix
multiplication on $\mathcal{H}_{\mathbb{PT}}$: $B$ and $C$ are in the
fundamental, rather than the adjoint. Similarly, $\overline{\partial}%
_{\mathbb{CP}^{1}}$ is a covariant derivative along the fuzzy $\mathbb{CP}%
^{1}$ fiber. (In the second term, some of the \textquotedblleft
bulk\textquotedblright\ form content of the $B$ and $C$ modes has been
absorbed into $\overline{\partial}_{\mathbb{CP}^{1}}$.) The derivatives
$\overline{\partial}_{\mathbb{PT}}$  and $\overline{\partial}_{\mathbb{CP}^{1}%
}$ correspond respectively to the vevs of the matrices $X_{k\times k}$ and $Y_{n\times n}$ in (\ref{xbcy}).

The parameters of the matrix model are the flux quanta $N$, and a continuous parameter $g_{\mathrm{MM}}$ as in eq. (\ref{gmm}).
Indeed, although the B-model is independent of K\"ahler data in the sense that we can always rescale $\Omega$, once we introduce
an explicit choice of flux and accompanying scale, this rescaling corresponds to adjusting a dilaton $g_{\mathrm{MM}}$.
The continuum limit is given by taking $N\rightarrow\infty$ with effective volume $\hslash^{3} \dim\mathcal{H}_{\mathbb{PT}}$
held fixed. At large $N$, perturbation theory of the matrix model organizes
according to the 't Hooft coupling $\lambda = g_{\mathrm{MM}}^{2}n N_{c}$.

\vspace{-3mm}

\subsection{{\protect\normalsize Amplitudes and Continuum Limit}}
\vspace{-3mm}

Natural observables of the twistor matrix model correspond to vevs involving
localized versions of the $B$ and $C$ modes. The analogue of maximally helicity violating (MHV) amplitudes are obtained by focussing on
the $B$, $C$ correlators on a single twistor line $\mathcal{H}_{x}$. Define
$B_{x}= B\cdot P_{x}$ and $C_{x}= P_{x}\cdot C$ with $P_{x}$ the projection
$\mathcal{H}_{\mathbb{PT}}\rightarrow\mathcal{H}_{x}$.
For normalized states $\left\vert z\right\rangle \in\mathcal{H}_{x}$, we can define the currents
\be
J_x(z) = \; C_{x}|z\rangle \langle z |\, B_{x}
\ee
and write the analogue of MHV amplitudes as a matrix model vev:
\be
A_{\rm MHV}(z_i) = \int\! d^{4|8}x\, \Bigl\langle {\rm Tr}\bigl(J_x(z_1) \ldots J_x(z_n)\bigr)\Bigr\rangle_{\mathrm{MM}}
\ee
where the trace is over $\mathcal{H}_x$ and the color indices.

$B_x$ and $C_x$ are maps from $\mathcal{H}_{\mathbb{CP}^{1}}$
to $\mathcal{H}_x$ and vice  versa. So they act as $n \times n$ matrices,
or equivalently,  degree $N$ homogeneous functions
of the projective coordinates $u,v,u^{\dag},v^{\dag}$ on a fuzzy $\mathbb{CP}^1 = \mathcal{H}_x$.
The kinetic operator $\overline{D}_{\mathbb{PT}}$ on the full twistor space, when sandwiched
between $B_x$ and $C_x$, thus naturally restricts to act only along the fiber.
Inserting $B=B_x$ and $C=C_x$, the matrix model action reduces to a trace over the
twistor line:%
\begin{equation}
\label{sred}
\mathcal{S}_{BC}\left(  B_{x},C_{x}\right)  =\text{Tr}_{\mathbb{CP}^{1}}%
(B_{x}\left(  \overline{\partial}_{\bar z} +\mathcal{A}_{\bar z} \right)
C_{x})
\end{equation}
where now the Dolbault operator $\overline{\partial}_{\bar z}$ is defined by a commutator with the holomorphic coordinates along the fuzzy sphere. All objects in eq. (\ref{sred}) act as $n\times n$ matrices. At large $N$, this action tends to that of a commutative $bc$ system
with action $\int d^{2} z \,\, b_{x} \overline{\partial}_{\mathcal{A}}|_{X} c_{x}$ on a
commutative twistor line $X$.  For states  $\left\vert z\right\rangle
,\left\vert w\right\rangle \in\mathcal{H}_{x}$ associated with points $z$ and
$w$ on $\mathbb{CP}^1$, the large $N$ vev of $\left\langle z|B_{x}%
C_{x}|w\right\rangle \sim\frac{1}{z-w} $.
So doing the $n\times n$ matrix integral over $B_x$ and $C_x$,
while taking the large $N$ limit, reproduces the correlators of the continuum
$bc$ system, which are known to generate the MHV amplitudes of ${\cal N}\!=\!4$ SYM theory \cite{Nair:1988bq}.

In the above argument, we assumed that the $B_x, C_x$ matrix integral neatly separates from
the modes at other locations along the $S^4$. We will now argue that this approximation is
justified at large $N$. The number of 
twistor lines $\mathcal{H}_{x_1},...,\mathcal{H}_{x_d}$ necessary to
span the bosonic part of $\mathcal{H}_{\mathbb{PT}}$ is $(N\!+\!1)(N\!+2)/2$.
Perform a decomposition $B\!=\!
{\sum}_iB_{i}\cdot P_{i}$ and $C\!=\!
{\sum}_j P_{j}\cdot C_{j}$, where
$P_{i}$ is the projection operator on the twistor line $\mathcal{H}_{x_i}$.
For sufficiently well separated $x_i$ and $x_j$, the action $S_{BC}(B_i,C_j)$
is highly suppressed at large $N$. The intuitive reason for this localization is that the $B$ and
$C$ modes are confined to small Landau orbits of angular size $1/\sqrt{N}$ (see eq. (\ref{overlap})),
due to the presence of the large background flux. Note that the kinetic operator $\overline{D}_{\mathbb{PT}}$ acts in the fundamental representation along the $S^4$, so all $B$ and $C$ modes are sensitive to the flux. So we see that for large $N$, the $B,C$ matrix model essentially
factorizes into a collection of decoupled $BC$ systems $\mathcal{S}_{BC}\left(  B_{i},C_{i}\right) $.

These $BC$ systems are still all coupled to the bulk field $\mathcal{A}$, which
propagates between the MHV\ vertices.
In the large $N$ limit this therefore reproduces the main features of the CSW
relations \cite{Cachazo:2004kj}, provided we
identify:
\begin{equation}
\lambda= g_{\mathrm{MM}}^{2}n N_{c} = g^{2}_{\mathrm{YM}} N_{c}.
\end{equation}
In this way, the matrix model naturally makes contact with the twistor gauge theory
proposed in \cite{Boels:2006ir}, based on the MHV generating function.

\vspace{-3mm}

\subsection{{\protect\normalsize Conclusion}}
\vspace{-3mm}

Fuzzy twistor space provides a novel covariant regulator for gauge theory.
Bulk space-time physics is dual to a large $N$ matrix model
obtained as a limit of holomorphic Chern-Simons theory in the
presence of a Yang monopole. The background flux introduces
an explicit length scale which demarcates the breakdown of UV
locality. Though the twistor matrix model shares many ingredients familiar from
the twistor string theory of \cite{Witten:2003nn}, we expect the
coupling to a closed string sector to evade the pathologies of
conformal gravity \cite{Berkovits:2004jj}. Rather, it is natural
to conjecture that the UV cutoff should be identified with
the Planck scale \cite{HVTO}.

\medskip

\textit{Acknowledgements:} We thank N. Arkani-Hamed, R. Boels, J. Broedel, S. Caron-Huot, D. Gross,
T. Hartman, J. Maldacena, E. Silverstein, D. Skinner, L. Susskind, E. Verlinde, E. Witten and M. Yamazaki
for helpful discussions. The work of JJH is supported by NSF grant
PHY-0969448. The work of HV is supported by NSF grant PHY-0756966.

\vspace{-3mm}


\begin{thebibliography}{99}                                                                                               %


\bibitem{PenroseRindler}R.~Penrose and W.~Rindler, \textit{Spinors and
Spacetime, Vols 1 \& 2}, Cambridge University Press (1986).

\bibitem{Witten:2003nn}E.~Witten,
Commun.\ Math.\ Phys.\ \textbf{252}, 189 (2004).

\bibitem{Heckman:2010pv}J.~J.~Heckman and H.~Verlinde,
JHEP \textbf{1101}, 044 (2011).

\bibitem{Penrose:1968me}R.~Penrose,
Int.\ J.\ Theor.\ Phys.\ \textbf{1}, 61 (1968).

\bibitem{Zhang:2001xs}S.-C.~Zhang and J.-P.~Hu,
Science \textbf{294}, 823 (2001). E.~Demler, S.-C.~Zhang,
Annals Phys.\  {\bf 271}, 83-119 (1999). For a stringy discussion see
M.~Fabinger,
JHEP \textbf{0205}, 037 (2002).

\bibitem{Yang:1977qv}C.~N.~Yang,
J.\ Math.\ Phys.\  {\bf 19}, 320 (1978);
J.\ Math.\ Phys.\  {\bf 19}, 2622 (1978).

\bibitem{Lechtenfeld:2005xi}O.~Lechtenfeld and C.~S\"{a}mann,
JHEP {\bf 0603}, 002 (2006).

\bibitem{Nair:1988bq}V.~P.~Nair,
Phys.\ Lett.\  B {\bf 214}, 215 (1988).

\bibitem{Cachazo:2004kj}F.~Cachazo, P.~Svrcek and E.~Witten,
JHEP {\bf 0409}, 006 (2004).

\bibitem{Mason:2010yk}L.~J.~Mason and D.~Skinner,
JHEP \textbf{1012}, 018 (2010).

\bibitem{Boels:2006ir}R.~Boels, L.~J.~Mason and D.~Skinner,
JHEP \textbf{0702}, 014 (2007).

\bibitem{HVTO}J.~J. Heckman and H.~Verlinde, \textit{to appear}.

\bibitem{Berkovits:2004jj}N.~Berkovits and E.~Witten,
JHEP {\bf 0408}, 009 (2004).

\end{thebibliography}
\end{document}